\documentclass[twocolumn]{aastex631}

\usepackage[flushleft]{threeparttable}
\usepackage{array,booktabs,makecell}
\usepackage{subfigure}
\usepackage{orcidlink}
\usepackage{tikz}
\usepackage{tablefootnote}
\usepackage{hyperref}
\usepackage{lineno}

\newcommand{\hcop}{HCO$^+$}
\newcommand{\dcop}{DCO$^+$}
\newcommand{\htcop}{H$^{13}$CO$^+$}

\newcommand{\nthp}{N$_2$H$^+$}

\newcommand{\solarmass}{M$_{\odot}$}
\def\farcs{\hbox{$.\!\!^{\prime\prime}$}} 

\shorttitle{DM Tau Ionization}
\graphicspath{{./}{figures/}}

\begin{document}

\title{Exploring the Complex Ionization Environment of the Turbulent DM Tau Disk}

\author{Deryl E. Long \orcidlink{0000-0003-3840-7490}} \affiliation{University of Virginia, 530 McCormick Rd., Charlottesville, VA 22903, USA}

\author{L. Ilsedore Cleeves \orcidlink{0000-0003-2076-8001}}
\affiliation{University of Virginia, 530 McCormick Rd., Charlottesville, VA 22903, USA}

\author{Fred C. Adams \orcidlink{0000-0002-8167-1767}}
\affiliation{Physics Department, University of Michigan, Ann Arbor, MI 48109, USA}

\author{Sean Andrews \orcidlink{0000-0003-2253-2270}}
\affiliation{Harvard-Smithsonian Center for Astrophysics, 60 Garden St., Cambridge, MA 02138, USA}

\author{Edwin A. Bergin \orcidlink{0000-0003-4179-6394}}
\affiliation{Department of Astronomy, University of Michigan, 1085 South University Avenue, Ann Arbor, MI 48109, USA}

\author{Viviana V. Guzmán \orcidlink{0000-0003-4784-3040}}
\affiliation{Instituto de Astrofísica, Pontificia Universidad Católica de Chile, Av. Vicũna Mackenna 4860, 7820436 Macul, Santiago, Chile}

\author{Jane Huang \orcidlink{0000-0001-6947-6072}}
\affiliation{Columbia University, 538 West 120th Street, New York, NY 10027, USA}

\author{A. Meredith Hughes \orcidlink{0000-0002-4803-6200}}
\affiliation{Van Vleck Observatory, Wesleyan University, 96 Foss Hill Dr., Middletown, CT 06459, USA}

\author{Chunhua Qi \orcidlink{0000-0001-8642-1786}}
\affiliation{Harvard-Smithsonian Center for Astrophysics, 60 Garden St., Cambridge, MA 02138, USA}

\author{Kamber Schwarz \orcidlink{0000-0002-6429-9457}}
\affiliation{Max Planck Institute for Astronomy, Königstuhl 17, Heidelberg, Germany}

\author{Jacob B. Simon \orcidlink{0000-0002-3771-8054}}
\affiliation{Department of Physics and Astronomy, Iowa State University, Ames, IA, 50010, USA}

\author{David Wilner \orcidlink{0000-0003-1526-7587}} \affiliation{Harvard-Smithsonian Center for Astrophysics, 60 Garden St., Cambridge, MA 02138, USA}

\correspondingauthor{Deryl E. Long}
\email{del6h@virginia.edu}



\begin{abstract}

Ionization drives important chemical and dynamical processes within protoplanetary disks, including the formation of organics and water in the cold midplane and the transportation of material via accretion and magneto-hydrodynamic (MHD) flows. Understanding these ionization-driven processes is crucial for understanding disk evolution and planet formation. We use new and archival ALMA observations of \hcop, \htcop, and \nthp\ to produce the first forward-modeled 2D ionization constraints for the DM Tau protoplanetary disk. We include ionization from multiple sources and explore the disk chemistry under a range of ionizing conditions. Abundances from our 2D chemical models are post-processed using non-LTE radiative transfer, visibility sampling, and imaging, and are compared directly to the observed radial emission profiles. The observations are best fit by a modestly reduced CR ionization rate ($\zeta_{CR}$ $\sim$ 10$^{-18}$ s$^{-1}$) and a hard X-ray spectrum (hardness ratio [HR] = 0.3), which we associate with stellar flaring conditions. Our best-fit model under-produces emission in the inner disk, suggesting that there may be an additional mechanism enhancing ionization in DM Tau's inner disk. Overall, our findings highlight the complexity of ionization in protoplanetary disks and the need for high resolution multi-line studies.

\end{abstract}

\keywords{Protoplanetary Disks --- Ionization --- Astrochemistry --- Planet Formation}


\section{Introduction} \label{sec:intro}

 \begin{figure*}
    \centering
    \includegraphics[scale=0.70]{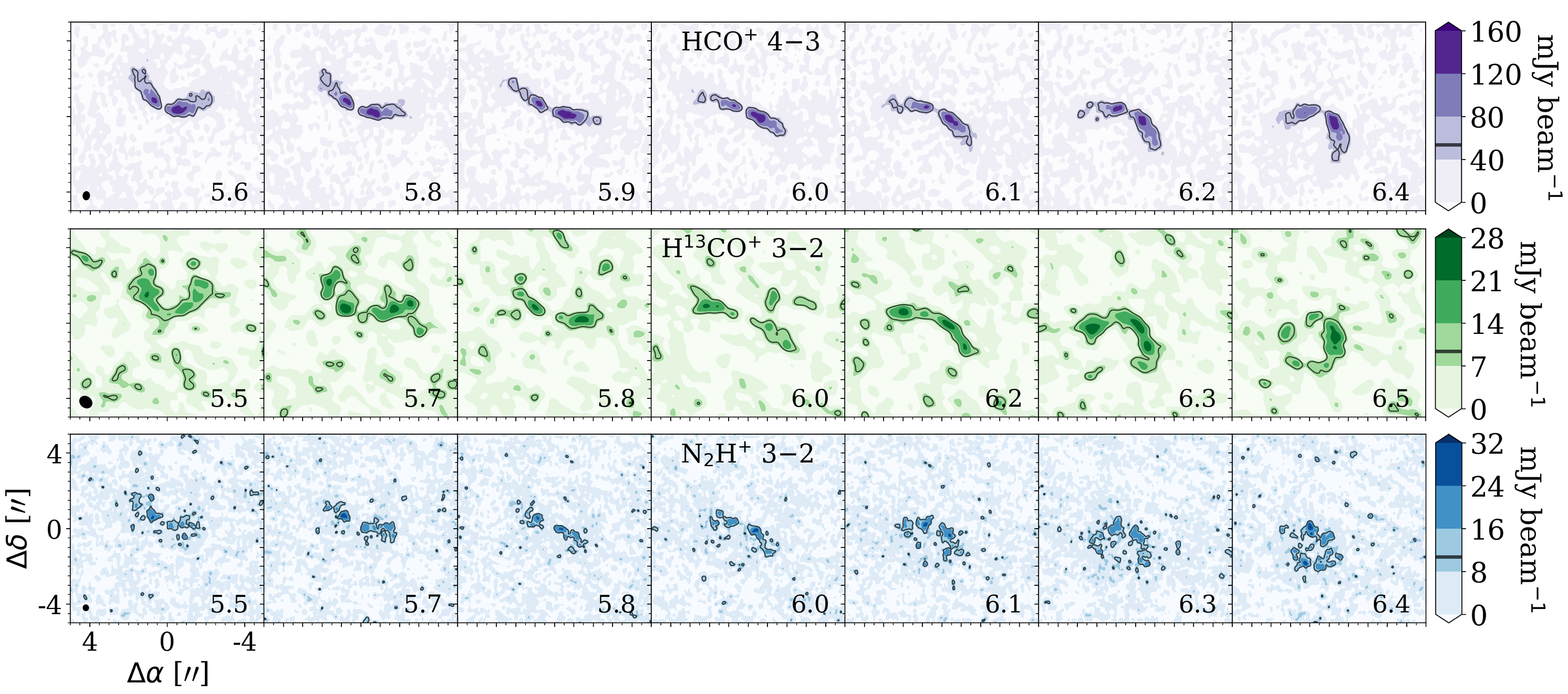}
    \caption{\normalsize{Channel maps for \hcop\ $J = 4 - 3$, \htcop\ $J = 3 - 2$, and \nthp \ $J = 3 - 2$. The solid contour represents 3$\sigma$. The beam is illustrated in the bottom left corner and the velocity relative to the local standard of rest in km s$^{-1}$ is indicated in the bottom right corner of each channel. }}
  \label{fig:chan}
\end{figure*}

Ionization plays a critical role in the evolution of protoplanetary disks and the formation of planets within these environments. Ions drive the most rapid chemical reactions \citep{1973ApJ...185..505H, 1982A&A...114..245T} in the coldest regions of the disk where neutral-neutral reactions are less efficient. Ionization also impacts the redistribution of planet-forming materials by driving transport via accretion and magneto-hydrodynamic (MHD) flows. Sufficient ionization allows the gas to couple to magnetic field lines and in turn drives a disk wind \citep{blandford_1982} and/or the magneto-rotational instability \citep[MRI;][]{balbus91}. Ionization is also a key input parameter for chemical models, the results of which are used to interpret observations of disks and in turn predict the compositions of forming planets \citep{eistrup_2018,price_2020}.

Ions are responsible for driving chemical complexity in the cold ($T<$ 50 K) disk midplane, creating formation pathways for organics \citep{cleeves16} and water \citep{vandishoeck2013}. If disks cannot efficiently form water, nascent planets would have to rely solely on inherited water from the parent molecular cloud \citep{cleeves_2014_water}. The efficiency of cold-water production may also vary with distance from the star, which would result in variations in ice across the disk. Understanding the distribution of ice is important, as the habitability of Earth-like planets may depend on post-formation delivery of water ice from asteroids and comets. The distribution of ice also influences the atmospheric and core compositions of gas giants \citep{oberg_2011, miotello_22}.

Ionization is also an important parameter for constraining hydro-- and magneto--hydrodynamic processes that impact the evolution of planet-forming material within the disk. In well-ionized regions of the disk, turbulence is thought to be primarily driven by the MRI. In weakly-ionized regions of the disk, turbulence may be present, but its strength and nature (e.g., degree of isotropy) will be modified by non-ideal effects, such as Ohmic diffusion \citep{flemming03,gole16}, ambipolar diffusion \citep{simon13a,simon13b}, and the Hall effect \citep{bai15,simon15b}. Moreover, recent modeling studies that consider non-ideal effects suggest that the accretion flow in the inner disk is largely laminar and that accretion heating is much weaker than previously assumed \citep{mori_19}. If ionization is low enough, then hydrodynamic instabilities such as the ``vertical shear instability" \citep[VSI;][]{nelson_2013}, ``convective overstability" \citep[COS;][]{klahr_2014}, and the ``zombie vortex instability" \citep[ZVI;][]{marcus_2015} affect the gas dynamics in a very different manner than do the magnetic effects (see e.g., \citealt{nelson_2013,fuksman23}). The gas flows driven by these various processes impact the conditions for planet formation. 

For planets to form, micron-sized dust grains must stick together to form aggregates and eventually planetesimals on size scales of $\sim$ 0.1--100 km. Several complications must be overcome in this process, including radial drift \citep{whipple_1972} and fragmentation \citep{blum_2018}. The ``streaming instability" is a promising mechanism for overcoming dust growth barriers, as dense collections of dust are able to form planetesimals via direct gravitational collapse \citep{youdin_2005}, although this mechanism can be suppressed by background turbulence if present (\citealt{Chen_2020,umurhan20,gole20,lim23}, but see \citealt{johansen07,yang18}). Thus the ionization state of the disk gas can influence the overall turbulence which can have attendant effects on planet formation.

 \begin{figure*}
    \centering
    \includegraphics[scale=0.65]{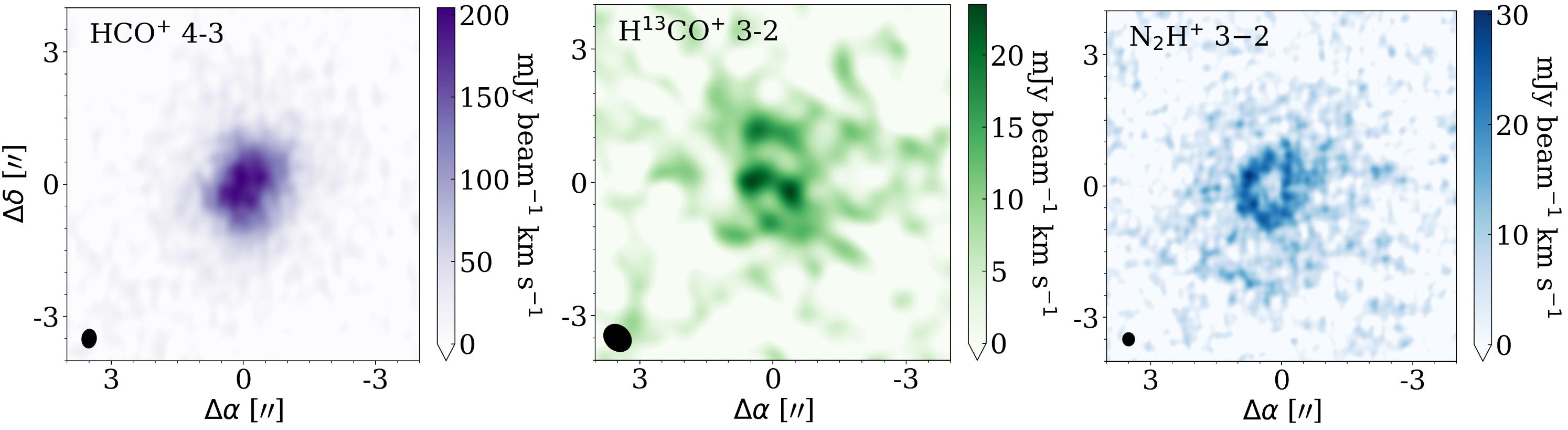}
    \caption{\normalsize{Integrated intensity (moment zero) maps for the resolved observations of \hcop\ $J = 4 - 3$, \htcop\ $J = 3 - 2$, and \nthp \ $J = 3 - 2$. The beam is illustrated in the bottom left corner. Moment zero maps were produced using the \texttt{bettermoments} \citep{teague_2018_bm} Python package.}}
  \label{fig:moments}\vspace{-0.5mm}
\end{figure*}

\begin{table*}
\caption{Line and Continuum Observations}
\begin{threeparttable}
\begin{tabular}{lccccc}
Transition  & Rest Freq. & Beam (PA) & Channel Width & RMS  & Integrated Flux$^{a}$ \\
  & (GHz) &  & (km s$^{-1}$) & (mJy beam$^{-1}$) & (Jy km s$^{-1}$)\\
     \midrule\midrule
\nthp \ $J = 4 - 3$ & 372.672  & 0\farcs66 $\times$ 0\farcs64 (7.0$^{\circ}$)  & 0.114 &  46  & 0.7 $\pm$ 0.24 \\ 
\hcop \ $J = 4 - 3$ &  356.734 & 0\farcs42 $\times$ 0\farcs32 (-6.2$^{\circ}$)  & 0.119  & 8.6  &  6.8 $\pm$ 0.69 \\ 
\nthp \ $J = 3 - 2$ & 279.512 & 0\farcs29 $\times$ 0\farcs27 (0.99$^{\circ}$) &  0.150  & 3.9  &  1.3 $\pm$ 0.14 \\
\htcop \ $J = 3 - 2$&  260.255 & 0\farcs67 $\times$ 0\farcs55 (46$^{\circ}$)  & 0.163 & 3.2  & 0.41 $\pm$ 0.046 \\ 
Continuum & 1 mm  & 0\farcs25 $\times$ 0\farcs16 (53$^{\circ}$)  & 1875 MHz$^{b}$ & 0.15 & 0.08 $\pm$ 0.008 Jy \\ 
Continuum & 1.3 mm  & 0\farcs38 $\times$ 0\farcs28 (-17$^{\circ}$)  & 1875 MHz & 0.27 & 0.16 $\pm$ 0.016 Jy \\ 
    \midrule\midrule
    \end{tabular}
    \begin{tablenotes}
\item[a] The disk integrated flux was measured within the same Keplerian masks used for imaging. Errors were estimated by applying the mask to line-free channels and include the nominal 10$\%$ flux calibration uncertainty from ALMA. \item[b] Bandwidth.
\end{tablenotes}
\end{threeparttable} 
\label{tab:obs}
\end{table*}

Ionization in disks is driven by several sources including UV photons, X-rays, and cosmic rays (CRs). Each of these sources ionize different vertical and radial regions of the disk. UV photons drive ionization at the disk surface but are quickly attenuated at depths greater than 10$^{-3}$ g cm$^{-2}$ \citep{bergin_2006_review}. X-rays are able to travel deeper into the disk, reaching depths from $\sim$ 10$^{-2}$ g cm$^{-2}$ for soft X-rays (1 keV) to 1.6 g cm$^{-2}$ for hard X-rays (10 keV). Thus, X-rays may have a significant impact in dense layers depending on the hardness of the stellar spectrum \citep{igea}. CRs are able to ionize gas column densities up to $\sim$ 100 g cm$^{-2}$, and are thus expected to be the most important sources of ionization in the most dense and cold regions of the disk where X-rays are strongly attenuated. 

The rate of CR ionization in the dense ISM is on the order of 10$^{-17}$ s$^{-1}$ but this rate could be reduced by orders of magnitude due to CR exclusion by stellar winds or magnetic fields \citep{cleeves13a,cleeves15_twhya} or enhanced due to local particle acceleration \citep{padovani_2015, padovani_2016}. Within the disk itself, detouring of CRs along sheared magnetic field lines can increase the effective column density by up to two orders of magnitude and reduce the rate of CR ionization in the disk midplane \citep{fujii_2022}. Stars forming in clusters may be exposed to even stronger external radiation via CRs as well as external X-rays and UV photons \citep{clarke_2007,adams_2010}. Recent observations of molecular ions in disks and protostars reveal evidence of spatial variation in ionization rates, possibly linked to the processes outlined above \citep{Seifert_2021, Aikawa_2021, cabedo_2023}. Ultimately, navigating the complex picture of ionization across the 2D disk environment requires a merging of detailed dynamical and chemical models with reliable observational constraints.

In this work we constrain the radially resolved ionization structure of the DM Tau protoplanetary disk using \hcop, \htcop, and \nthp. DM Tau is a large \citep[gas disk radius $\sim$ 800 au;][]{dartois_2003}, well studied disk around a 0.54 \solarmass \ T Tauri star \citep{Simon_2000}. DM Tau is one of only two disks in which turbulence has been measured \citep{flaherty, flaherty_24}, and it has yet to be determined whether this turbulence is ionization-driven via magnetic effects, such as the MRI. Previous work suggests that DM Tau's disk is well--ionized in the warm molecular layer \citep[$x_{e} \sim$ 10$^{-7}$;][]{Teague_2015} with the ionization fraction decreasing toward the midplane \citep{oberg11_dmtau}, consistent with theoretical expectations. We present new ALMA observations of molecular ions in DM Tau and use an updated physical model and 2D chemical model to explore the role of different ionization sources (namely CRs and X-rays) in driving ionization in DM Tau. We also discuss the usefulness of the different molecular lines for studying ionization in this disk and what they tell us, individually and in concert.

 \begin{figure*}
    \centering
    \includegraphics[scale=0.70]{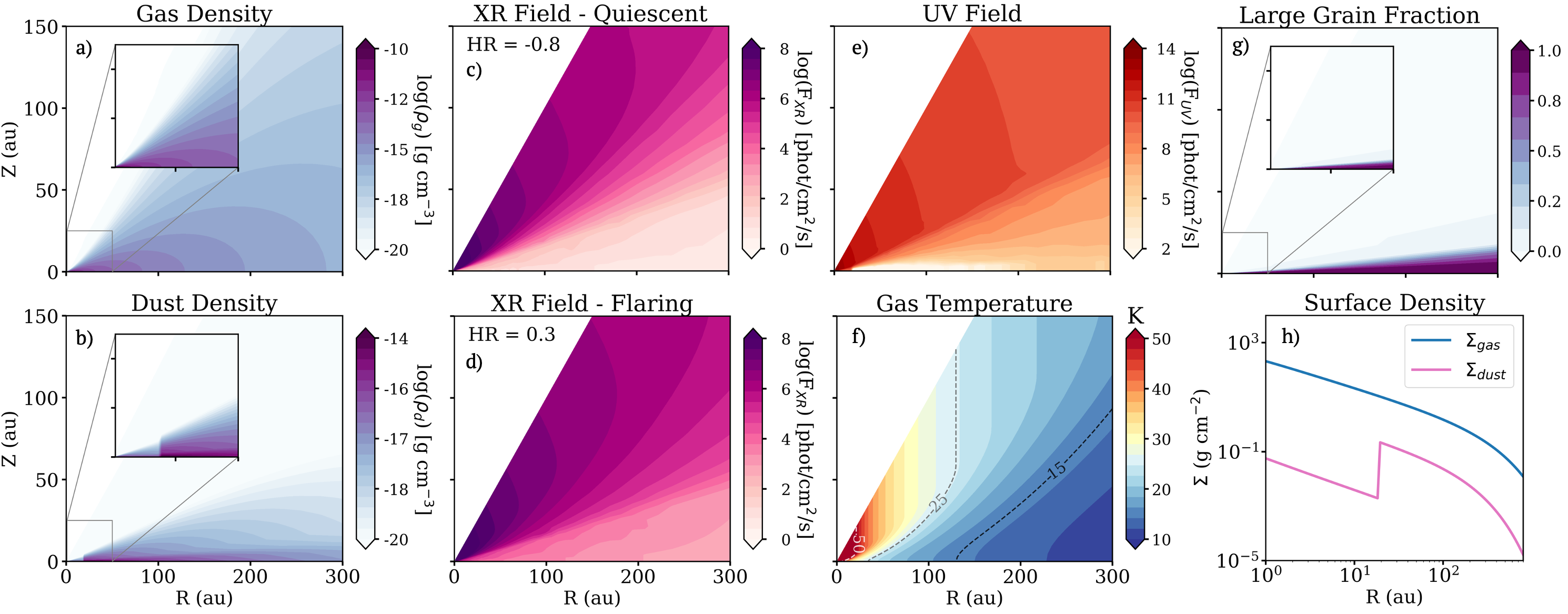}
    \caption{\normalsize{Panel a) shows the gas densities from \citet{flaherty} and panel b) shows the total dust density including both small and large grains. Panels c) and d) show the integrated X-ray flux from 1-20 keV for the quiescent and flaring X-ray models. Panel e) shows the wavelength integrated UV flux and panel f) shows the gas temperature structure with 50 K, 25 K, and 15 K contours. Panel g) shows the fraction of large grains in the disk, which are concentrated in the disk midplane. Panel h) shows the surface density profiles for our gas and dust models.}}
  \label{fig:structure_all}
\end{figure*}

\section{ALMA Observations}\label{sec:obs}
Observations of \hcop \ $J = 4 - 3$, \nthp \ $J = 4 - 3$, and \htcop \ $J = 3 - 2$ were taken with ALMA under project code 2019.1.00379.S (PI: Cleeves). We also made use of data from project 2015.1.00678.S (PI: Qi) to constrain \nthp \ $J = 3 - 2$. A detailed description of those observations and their calibration can be found in \citet{qi_2019}. The Band 7 \hcop\ $J = 4 - 3$ and \nthp\ $J = 4 - 3$ data were taken on 2019 October 17-18 with 43 antennas covering 15m-740m baselines. These data were calibrated with the CASA Pipeline, with J0423-0120 as the band-pass calibrator and J0510+1800 as the phase calibrator. The Band 6 \htcop\ $J = 3 - 2$ data were taken on 2021 July 7 with 43 antennas covering 14m-2492m baselines.

For the \htcop \ data set we saw no improvement in the continuum RMS with phase self-calibration, so the data were calibrated only with the standard pipeline and subsequently imaged. For the data set with \hcop \ $J = 4 - 3$ and \nthp \ $J = 4 - 3$ we applied one round of phase self-calibration which resulted in a small factor of 1.1x improvement in the RMS of the continuum data. We performed continuum subtraction and imaging for all lines using the \texttt{uvcontsub} and \texttt{tclean} commands from CASA \texttt{v5.6.1} and Keplerian masks generated with the \texttt{keplerian\textunderscore mask}\footnote{\url{https://github.com/richteague/keplerian_mask}} Python package \citep{teague_2020_kep}. The masks for both \nthp \ lines contain three components to capture the hyperfine emission structures. For the line imaging of \hcop \ we used Briggs weighting and a robust parameter of 0.5. For \htcop \ and \nthp \ $J = 4 - 3$ we used natural weighting and a \texttt{uvtaper} of 0\farcs5 to maximize sensitivity. Following \citet{qi_2019} we imaged \nthp \ $J = 3 - 2$ using natural weighting.

The data for \nthp \ $J = 4 - 3$ was significantly more noisy than the other lines due to its close proximity to atmospheric water vapour features. We used the \texttt{matched\textunderscore filter} command from the \texttt{VISIBLE}\footnote{\url{https://github.com/AstroChem/VISIBLE}} Python package \citep{Loomis_2018} with our \hcop \ detection as a filter. We saw a 6$\sigma$ response, indicating the presence of \nthp \ $J = 4 - 3$. Though the \nthp $J = 4 - 3$ line is not resolved in the observations we still report the integrated flux and include this line in our analysis. Channel maps for the three resolved line detections are shown in Figure \ref{fig:chan} and additional information about the line and continuum observations is listed in Table \ref{tab:obs}. We create velocity integrated intensity (zeroth moment) maps (Figure \ref{fig:moments}) for the resolved detections using the \texttt{bettermoments}\footnote{\url{https://github.com/richteague/bettermoments}} Python package \citep{teague_2018_bm}. We report the RMS measured in line-free channels. We produce radial line intensity profiles using the \texttt{GoFish}\footnote{\url{https://github.com/richteague/gofish}} Python package \citep{GoFish} and discuss these in more detail in Section \ref{sec:res}.

\section{Modeling DM Tau}\label{sec:mod}
\subsection{Physical Model}\label{sec:mod-phys}

Our physical model combines a well-constrained gas density and temperature structure from \citet{flaherty} with a dust structure based on continuum visibilities and spectral energy distribution (SED) fitting presented in \citet{Andrews_2011}. The gas density and temperature structures were parameterized and fit using the Markov Chain Monte Carlo (MCMC) routine
EMCEE \citep{foreman_mackey} and high resolution CO observations. A detailed description of the CO modeling and temperature parametrization can be found in \citet{flaherty}. The dust structure is derived from the azimuthally symmetric gas surface density profile

\begin{equation}
    \Sigma_{g} = \Sigma_{c} \left(\frac{R}{R_{c}}\right)^{-\gamma} \exp \left[ - \left(\frac{R}{R_{c}}\right)^{2 -\gamma} \right],
\end{equation}

 \noindent where the characteristic scaling radius $R_c$ = 135 au was determined by \citet{Andrews_2011} and the surface density gradient is set to $\gamma$ = 1. Our model assumes a global dust-to-gas ratio of $\xi$ = 0.01. The dust disk has a flared vertical structure with a scale height distribution given by

 \begin{equation}
     H(r) = H_{100}\left(\frac{r}{\text{100 au}}\right)^{\psi} 
 \end{equation}
 
 \noindent where $H_{100}$ = 9.75 au and $\psi$ = 1.2. Following the SED fitting results in \citet{Andrews_2011}, we account for the disk's inner gap by depleting the dust by a factor $\delta_{cav}$ = 4.8 in the region $R < R_{cav}$ = 19 au. With the resulting surface density profile we calculate dust densities for small (0.005 $\mu$m -- 1 $\mu$m) and large (0.005 $\mu$m -- 1 mm) grains using the density distributions 
 
 \begin{equation}
     \rho_{s} = \frac{(1-f)\Sigma_{d}}{\sqrt{2\pi}H}\exp\left[-\frac{1}{2}\left(\frac{z}{H}\right)^{2}\right]
 \end{equation}
 
\noindent and

\begin{equation}
    \rho_{l} = \frac{f\Sigma_{d}}{\sqrt{2\pi}H \chi}\exp\left[-\frac{1}{2}\left(\frac{z}{\chi H}\right)^{2}\right] \,,
\end{equation}

\noindent where \textit{f} and $\chi$ are settling parameters fixed at 0.85 and 0.2 respectively, meaning that large grains make up 85$\%$ of the total dust column and are distributed up to 20$\%$ of the disk scale height. Both the small and large dust grain populations follow an MRN size distribution \citep{Mathis_77} within their respective size ranges. The distribution of gas, large dust grains, small dust grains, and temperature structure are shown in Figure \ref{fig:structure_all}. 

\subsection{Ionization Model}\label{sec:mod-ion}
Our models include ionization from three sources: UV photons, X-ray photons, and CRs. We exclude effects of inherited short-lived radionuclides (SLRs) for now since their initial abundance is uncertain and any inherited SLRs' rate of ionization will have decayed over DM Tau's estimated lifetime of $\sim$ 3--7 Myr \citep{Simon_2000}, and ultimately should have negligible contributions to bulk disk ionization at the disk's current age.

 \begin{table}[]
  \caption{CR--ionization rates ($\zeta_{CR}$) used for our modeling suite. Here ``Min Mod" and ``Max Mod" refer to the efficiency of modulation by stellar winds \citep{cleeves13a}. In addition to the five $\zeta_{CR}$ values listed here, we also model two different X-ray conditions, yielding a modeling suite with a total of 10 models.}
    \vspace{-0.5mm}
  \centering
\begin{tabular}{lclclc}
\midrule\midrule  
Cosmic Ray Model  & Label &  $\zeta_{CR}$ \\
\hline
\citet{mosk_2002} & M02 & 6.8 $\times$ 10$^{-16}$ \\
\citet{webber_1998} & W98 & 2.0 $\times$ 10$^{-17}$ \\
Solar System Min Mod & SSM & 1.1 $\times$ 10$^{-18}$ \\ 
Solar System Max Mod & SSX & 1.6 $\times$ 10$^{-19}$ \\
T Tauri Max Mod & TTX & 1.0 $\times$ 10$^{-21}$ \\
    \midrule\midrule
    \end{tabular}
\label{tab:cr}
  \end{table}


Two-dimensional stellar X-ray and UV fluxes (Figure \ref{fig:structure_all}) were obtained using the \citet{bethell_bergin} Monte Carlo radiative transfer code and cross sections. We use an observed stellar UV spectrum for TW Hya and scale it to DM Tau's FUV luminosity by a factor of 2.16. We explore the effects of both quiescent and ``flaring" X-ray states using template spectra from \citet{cleeves15_twhya}. Both spectra are scaled to an X-ray luminosity of 2.5 $ \times$ 10$^{29}$ erg s$^{-1}$ for DM Tau, which is a factor of $\sim$ 2 below the upper limit of 4.6 $ \times$ 10$^{29}$ erg s$^{-1}$ from \citet{damiani_1995} after excluding energies $<$ 1 keV. Our X-ray models include energies from 1--20 keV. The quiescent and flaring states are distinguished by their respective hardness ratios, HR = ($L_{soft} - L_{hard})/(L_{soft} + L_{hard}$). The quiescent model (HR = -0.8) is soft X-ray dominated, while the hardened ``flaring" model (HR = 0.3) is hard X-ray dominated. Panels \textit{c} and \textit{d} in Figure \ref{fig:structure_all} demonstrate the resulting difference in flux between the two XR models. 

   \begin{table}[]
  \caption{Model Initial Abundances relative to total Hydrogen density. The (gr) suffix indicates that the species is initialized on grains instead of in the gas phase. The elemental abundances for dominant molecules N, C, and O are 7.5$\times$ 10$^{-5}$, 5$\times$ 10$^{-5}$, and 2.5$\times$ 10$^{-4}$ respectively.}
    \vspace{-0.5mm}
  \centering
\begin{tabular}{lclc}
\midrule\midrule  
Molecule  & Abundance & Molecule & Abundance  \\
\hline
H$_{2}$ & 5.00 $\times$ 10$^{-1}$ & He & 1.40 $\times$ 10$^{-1}$ \\
N$_{2}$ & 3.75 $\times$ 10$^{-5}$ & HCN & 1.00 $\times$ 10$^{-8}$ \\
CS & 4.00 $\times$ 10$^{-9}$ & SO & 5.00 $\times$ 10$^{-9}$ \\
\hcop & 9.00 $\times$ 10$^{-9}$ & H$_{3}$$^{+}$ & 1.00 $\times$ 10$^{-8}$ \\
C$_{2}$H & 8.00 $\times$ 10$^{-9}$ & CO & 2.60 $\times$ 10$^{-5}$ \\
H$_{2}$O(gr) & 1.00 $\times$ 10$^{-4}$ & Si$^{+}$ & 1.00 $\times$ 10$^{-11}$ \\
Mg$^{+}$ & 1.00 $\times$ 10$^{-11}$ & Fe$^{+}$ & 1.00 $\times$ 10$^{-11}$ \\
Grains & 6.00 $\times$ 10$^{-12}$ & & \\
    \midrule\midrule
    \end{tabular}
\label{tab:abunds}
  \end{table}

Our modeling suite includes five different incident CR--ionization rates ($\zeta_{CR}$) listed in Table \ref{tab:cr} and CR attenuation as a function of column density, as described in \citet{cleeves13a}. Our two highest CR rates, M02 and W98, represent two different estimates of the local interstellar (LIS) cosmic ray spectrum. The \citet{mosk_2002} spectrum is the best-fit model based on the diffuse $\gamma$-ray background as well as other data including the proton, antiproton, and alpha particle spectra. The rate from \citet{webber_1998} approximates the conditions of the dense ISM based on measurements taken by \textit{Voyager} and \textit{Pioneer}. The Solar System Minimum-Modulation (SSM) and Maximum-Modulation (SSX) are based on measured CR rates on Earth, representing the CR ionization for our solar system given minimum and maximum amounts of CR modulation via stellar winds. Our lowest CR rate is the T Tauri Maximum-Modulation (TTX), which was extrapolated by \citet{cleeves13a} to account for enhanced CR modulation by the higher stellar wind outflow rates measured for T Tauri stars. Additional details for the CR rates can be found in \citet{cleeves13a}. 

\begin{figure}
    \centering
    \includegraphics[width=\columnwidth]{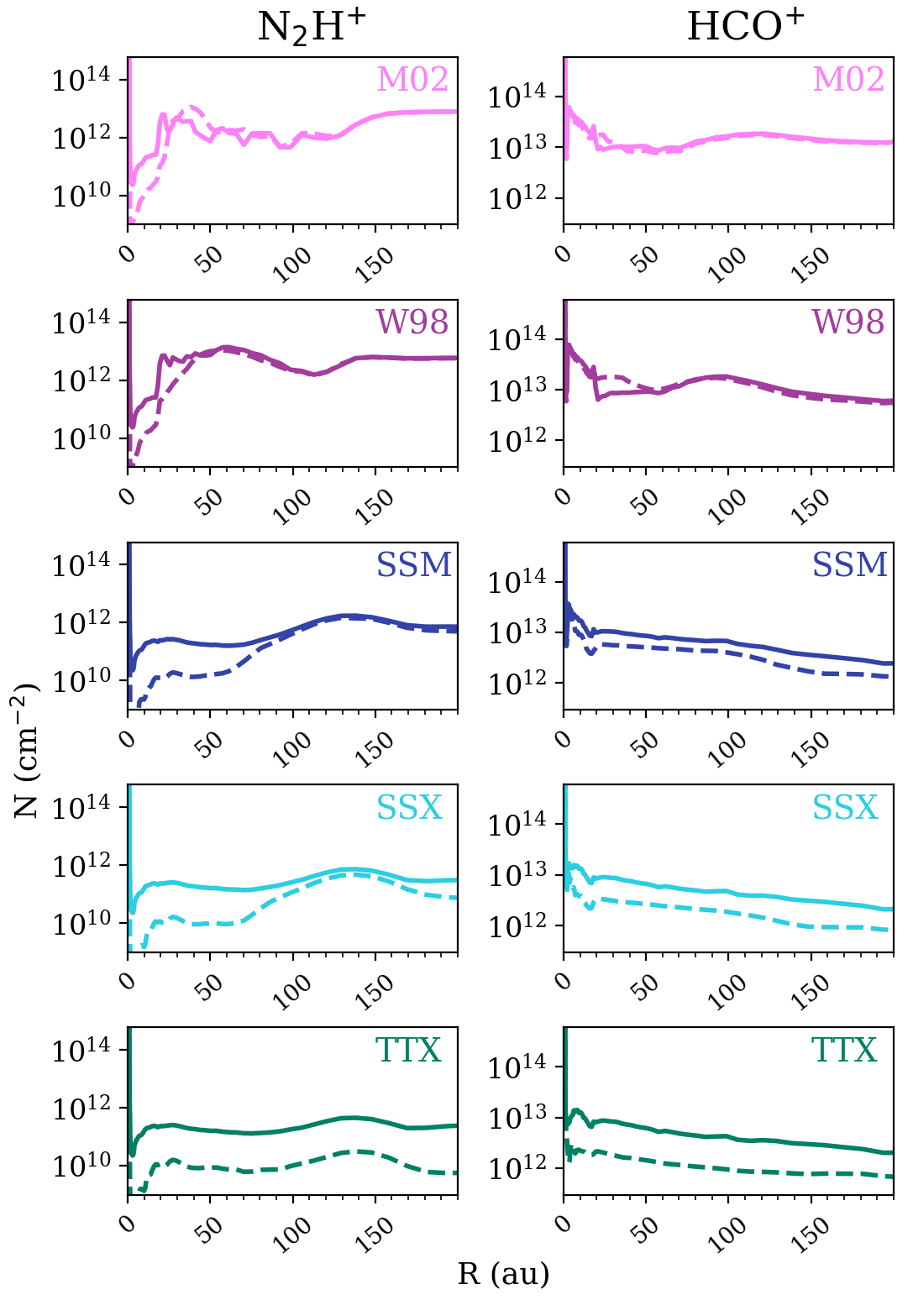}
    \caption{\normalsize{Vertically integrated radial column densities for \nthp \ (left) and \hcop \ (right). The cosmic ray ionization rate decreases from top to bottom with labels corresponding to the rates listed in Table \ref{tab:cr}. Models with quiescent X-rays are shown with dashed lines while models with hardened X-rays are solid lines.}}
  \label{fig:colden}
\end{figure}

\subsection{Chemical Model}\label{sec:mod-chem}

 Chemical abundances are calculated as a function of position and time using a 2D time--dependent rate equation chemical code from \citet{fogel_2011} adapted in \citet{cleeves_2014_water} and \citet{anderson_2021}. The results presented in this work utilize our latest updated chemical network with 18,608 reactions and 1038 species, including deuterium isotopologues. Our network does not include carbon isotopes $^{12}$C and $^{13}$C so we retrieve \htcop \ abundances by applying a constant $^{12}$C/$^{13}$C ratio of 60 \citep{langer_penzias_1993}, however in reality we expect that the gas-phase $^{12}$C/$^{13}$C ratio may have radial variations, particularly across the CO snowline \citep{yoshida_22}. Our initial abundances (Table \ref{tab:abunds}) are representative of the chemical conditions in a typical molecular cloud. However, it has been shown that DM Tau's CO depletion ranges from a factor of 1-10 radially across the disk \citep{Zhang_2019}. Based on this range we choose to use an ``average" CO depletion factor of 5 such that our starting CO abundance is 2.6 $\times$ 10$^{-5}$. We do not aim to determine if the CO destruction in DM Tau occurs within the lifetime of the disk or prior \citep[e.g.,][]{bergner_2020}, but rather aim to create a chemical environment reflective of DM Tau's present state.

 We run a grid of models using the physical structure and radiation fields shown in Figure \ref{fig:structure_all}. With five different CR ionization rates and two XR fields we have a total of ten models. Models are computed out to a time of 1 Myr, at which point the ion chemistry has settled into a pseudo-steady state. Figure \ref{fig:colden} shows the resulting model column densities for \nthp\ and \hcop.

 \begin{figure}
    \centering
    \includegraphics[width=\columnwidth]{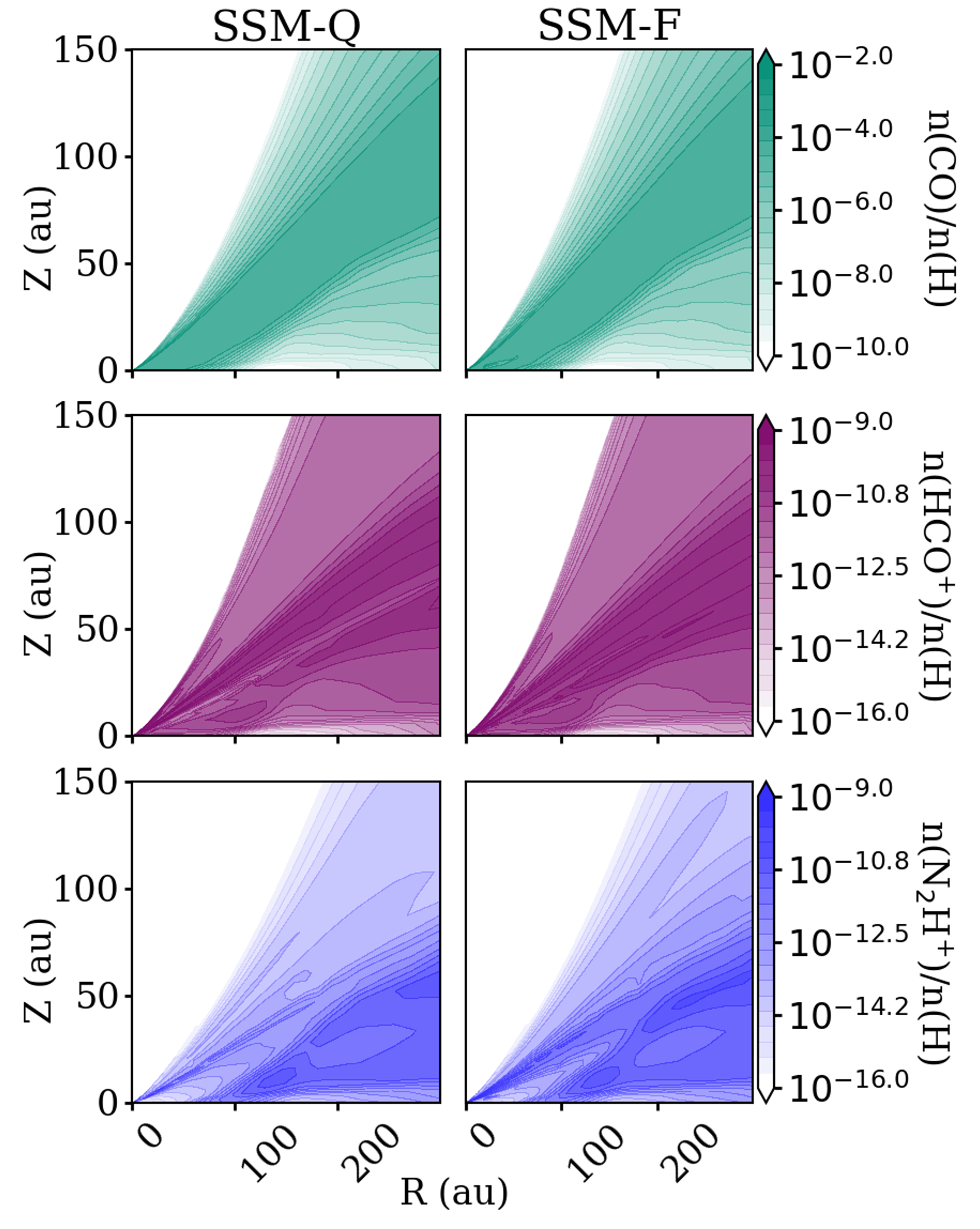}
    \caption{\normalsize{Comparison of SSM model abundances under quiescent and flaring X-ray conditions after 1 Myr for CO (top), \hcop\ (middle), and \nthp\ (bottom).}}
  \label{fig:abunds}
\end{figure}

 \begin{figure*}
    \centering
    \includegraphics[scale=0.78]{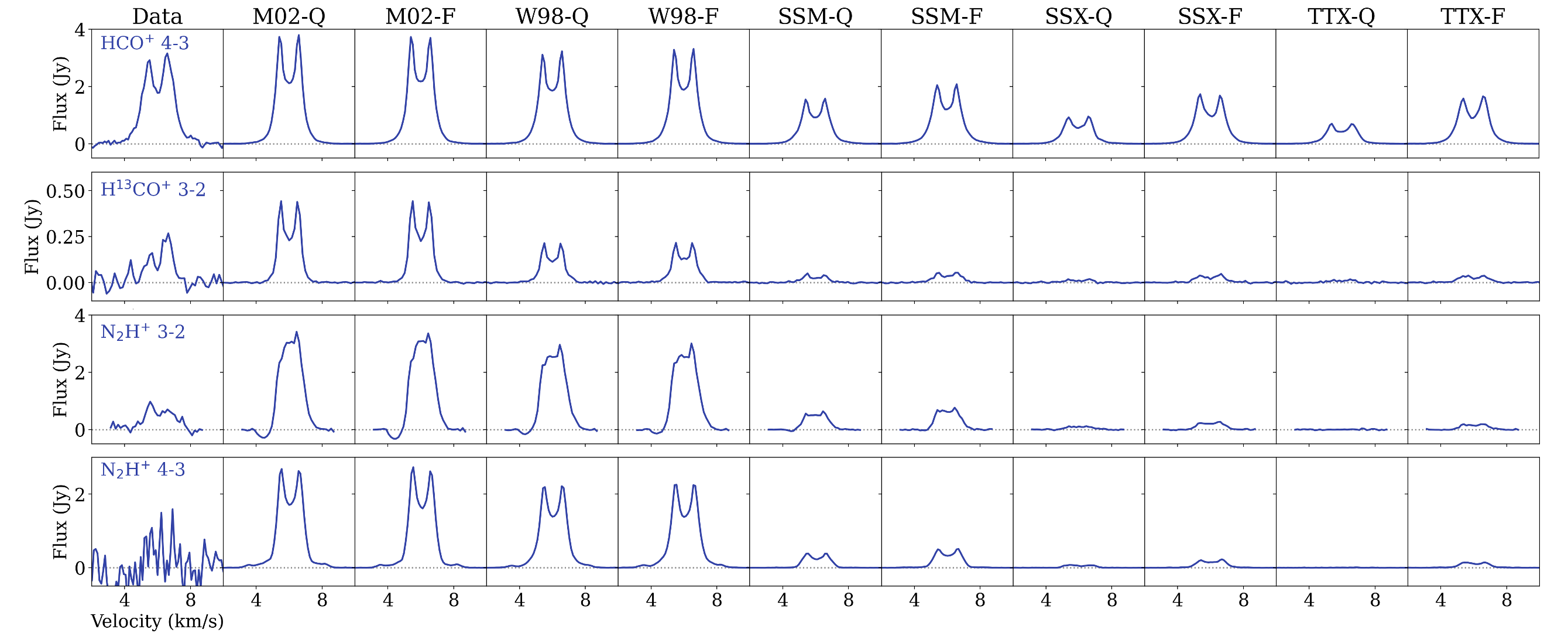}
    \caption{\normalsize{Observed and model spectra for the standard turbulence models produced using \texttt{GoFish}. }}
  \label{fig:spec}\vspace{1mm}
\end{figure*}

 For \nthp, increasing the CR ionization rate ($\zeta_{CR}$) tends to increase the column density and modulates the shape of the radial column density profile. Increasing the hardness of the X-rays drives the column densities up, particularly in the inner disk. This effect becomes more pronounced throughout the disk as $\zeta_{CR}$ decreases.

 For \hcop \ we also see a general trend of increasing column density with increasing $\zeta_{CR}$. For all of the models we see a drop in \hcop \ column density at the location of the dust wall at 20 au. This corresponds to increases in \nthp \ column density, though this is much less noticeable in the low CR models. We expect to see this behavior because these two species are opposed in their formation pathways: \nthp \ thrives in the absence of gas-phase CO, while \hcop\ requires the gas-phase CO in order to form. This relationship between the two species is most pronounced in the models with high $\zeta_{CR}$ and hard X-rays, suggesting that the formation of \nthp \ and \hcop \ in the high CR models is strongly influenced by X-ray related chemistry in the inner disk in the presence of a dust gap. We discuss the potential role of DM Tau's substructure in modulating ionization chemistry in more detail in Section \ref{sec:disc}.

Figure \ref{fig:abunds} shows the chemical abundances of CO, \hcop, and \nthp\ for a quiescent X-ray state and a flaring X-ray state. Similar to what we see in the model column densities, both \hcop\ and \nthp\ abundances appear to increase slightly in certain regions of the disk for the case of the hard X-ray spectrum, though the differences in abundance are minor. For example, there appears to be an enhancement of both \nthp\ and \hcop\ at z/r $\sim$ 0.2.

\subsection{Synthetic Observation Comparison}\label{sec:mod-comp}

The resulting disk abundances are input to the \texttt{LIME} non-LTE radiative transfer code \citep{brinch_2010} in order to calculate the emergent flux and optical depth for \hcop, \nthp, and \htcop. We use collisional rates from the LAMDA database \citep{schoer_2005,daniel_2005,denis_2020}. The \texttt{LIME} code includes a doppler broadening parameter (b) that accounts for line broadening due to local turbulence. We set the doppler velocity as a function of the local sound speed (c$_{s}$), which is temperature dependent. For our ``standard turbulence" models we use b = 0.2c$_{s}$. This degree of turbulence is slightly lower than the conservative turbulence estimate (0.25c$_{s}$) from \citet{flaherty}. We also run \texttt{LIME} models with high turbulent velocities (b = 0.5c$_{s}$) for comparison.

The \texttt{LIME} models for \hcop, \htcop, and \nthp \ are spectrally over-sampled and channel averaged by a factor of 20 to mimic channel smearing seen in real observations. The resulting fits cubes have the same number of channels and the same channel widths as the observations listed in Table \ref{tab:obs}. All of the models assume a distance of 144.5 pc \citep{lindegren_2018} and an inclination of 35.2$^{\circ}$ \citep{Kudo_2018}. 

The optical depth profiles (Figure \ref{fig:tau_profs}, Appendix A) show that \hcop \ tends to be optically thick in the inner disk as expected, while \htcop\ is optically thin throughout the disk for all models. We find that \hcop \ appears to become more optically thin towards the inner disk ($<$ 200 au) for low CR and/or quiescent XR models. Both transitions of \nthp\ are generally optically thin throughout the disk, but become more optically thick when column densities exceed  $\sim$ 10$^{12}$ cm$^{-2}$ (this is only the case for the high CR models). Similar to the column density profiles, we observe radial variations in the optical depth profiles corresponding to the dust gap at 20 au. Additionally, the \nthp\ optical depth profiles exhibit a peak between 150-200 au for all models, and an additional peak around 50 au that is most pronounced in the high CR models.

We use \texttt{vis\textunderscore sample} \citep{Loomis_2018} to perform visibility sampling of our LIME fits cubes. Figure \ref{fig:spec} shows the data spectra compared to the model spectra for the standard turbulence models. By using the ALMA measurement sets from our observations as inputs, we ensure that our simulated observations have the same uv coverage as the real observations. We perform continuum subtraction and image the model measurement sets in CASA \texttt{v.5.6.0} with the same parameters described in Section 2. To make direct comparisons in the image plane, we produce radial profiles of the observed and modeled emission after imaging. Radial profiles are generated using the \texttt{radial\textunderscore profile} function in the \texttt{GoFish} program \citep{GoFish}. We apply the same Keplerian masks used to image the observations and integrate emission over the same velocity ranges used to produce the moment zero maps shown in Figure \ref{fig:moments}.

 \begin{figure*}
    \centering
    \includegraphics[scale=0.65]{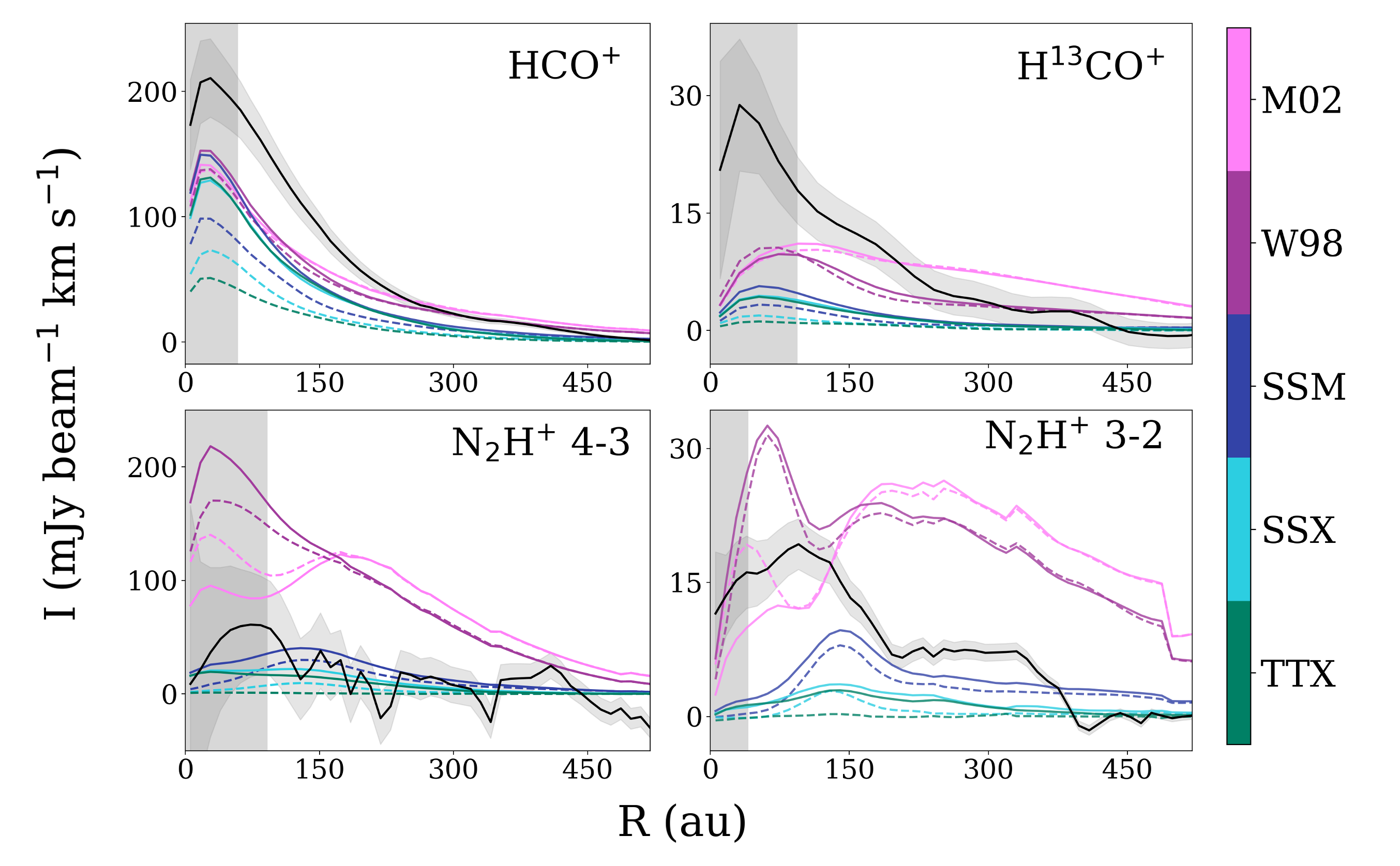}
    \caption{\normalsize{Observed and model radial profiles for the full disk. In each panel the solid black line is the observed emission profile and the associated light gray shading represents the uncertainties on the profile from \texttt{GoFish} plus the nominal 10\% flux uncertainty from ALMA. The five different CR models are color coded according to the colorbar and labels on the right. For the models, dashed lines denote a quiescent XR spectrum while solid lines denote a hardened XR spectrum. The dark grey shading represents the beam of the data in au.}}
  \label{fig:radprofs_glob}\vspace{1mm}
\end{figure*}

\section{Results}\label{sec:res}

We measure the goodness of fit for each model by calculating the reduced $\chi^{2}$ values for the model radial profiles of \hcop, \htcop, and \nthp. We perform global fitting for standard and high turbulence models (Section \ref{sec:res-glob}) as well as outer disk fitting for standard and high turbulence models (Section \ref{sec:res-out}). The decision to work in the image plane was motivated by previous findings that the ionization profile of IM Lup showed a radial gradient \citep{Seifert_2021}. We choose to work in the image plane so that all radii are weighted equally in our fitting routine, as opposed to the visibility plane where it is possible for brighter visibilities at long baselines to dominate the fitting results. Thus by working in the image plane and with the radial profiles, we can more easily examine where our models are in agreement and identify ionization gradients if present. We use the radial profile reduced $\chi^{2}$ values as a guide for identifying the best-fit models. While high reduced $\chi^{2}$ values are expected due to the difficulty of reproducing emission across the full extent of the disk, this analysis ultimately provides a strong constraint on the best-fit model relative to the rest of the model grid.

 \begin{figure*}
    \centering
    \includegraphics[scale=0.73]{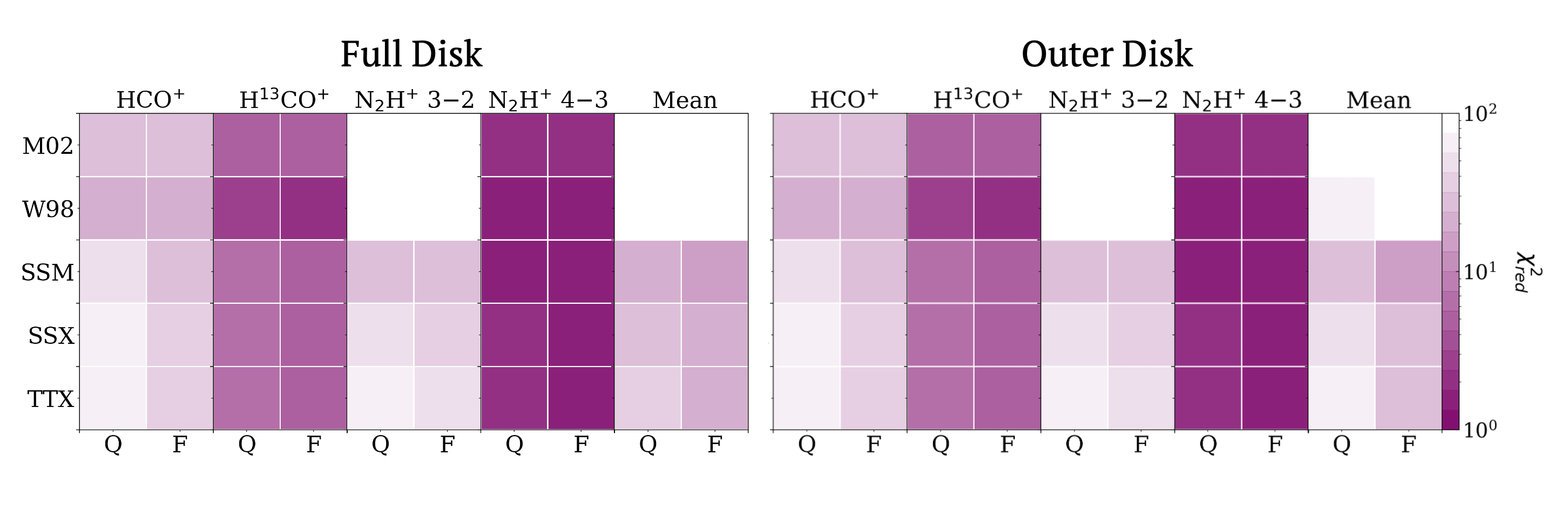}
    \caption{\normalsize{Reduced $\chi^{2}$ fitting results for both the full disk and outer disk ($>$ 200 au). The cosmic ray ionization rates, $\zeta_{CR}$, are labeled on the left from high to low. On the x-axis, Q denotes a quiescent XR model while F denotes a flaring XR model. For plotting purposes we discard extremely bad fits by cutting off reduced $\chi^{2}$ values above 100, but these values still contribute to the ``Mean" $\chi^{2}_{red}$ scores.}}
  \label{fig:chisq_all}\vspace{-1mm}
\end{figure*}

\subsection{Global Model Fitting}\label{sec:res-glob}

Observed emission reaches zero for all four lines by 500 au, so we consider radii $<$ 500 au when calculating the global goodness of fit. Observed and model radial profiles for the full disk are shown in Figure \ref{fig:radprofs_glob} and the reduced $\chi^{2}$ fitting results are shown in Figure \ref{fig:chisq_all}. 

\subsubsection{Standard Turbulence Models}\label{sec:res-glob-st}
We first simulate the model emission with a disk of standard turbulence (b = 0.2c$_{s}$). We find that the \hcop\ and \htcop\ tend to prefer a W98-F (high ionization rate) model, while in contrast, the \nthp\ lines strongly prefer a SSM-F (reduced ionization rate) model.  

The observed \hcop\ and \htcop\ emission profiles each exhibit a small inner deficit and emission peaks at 30 and 40 au respectively. The models follow similar behavior, showing little variation in the location of the emission peak. The observed inner deficits in \hcop\ and \htcop\ emission may be related to the efficient destruction of \hcop\ by H$_{2}$O in the inner disk \citep{leemker_2021,notsu_2021}, however we expect the shapes of the emission rings seen here to be dominated by continuum opacity and excitation effects. Models with hard XR spectra have brighter peaks than their quiescent counterparts. The disparity between the hard vs. quiescent model emission peaks is wider for low CR models, suggesting that hard X-ray spectra have more of an impact in low $\zeta_{CR}$ environments ($\lesssim$ 10$^{-18}$ s$^{-1}$), as expected, since X-rays can become competitive in the absence of CRs. Since the models are all under-bright in the inner disk, a model that produces a bright emission peak (high $\zeta_{CR}$ and hardened XR spectrum) is preferred, and W98-F emerges as the best global fit for both \hcop \ ($\chi^{2}_{red}$ = 19.3) and \htcop \ ($\chi^{2}_{red}$ = 2.3).

The observed \nthp\ $J = 3 - 2$ emission profile exhibits a peak at 100 au and an emission ``plateau" between $\sim$ 200-350 au followed by a stark drop-off. The observed profile for \nthp\ $J = 4 - 3$ is very noisy due to poor atmospheric transmission at this line's frequency, but it shows a peak around 80 au. We opt to include this line in our radial profile fitting because it demonstrates that the high-CR models are over-bright, but we do not gain much insight from the emission morphology due to its being unresolved. 

However, the 3--2 observations are well resolved, and we can see that our models struggle to reproduce the observed morphology. Low CR models are under-bright in the inner disk and peak further out, while high CR models are extremely over-bright in the outer disk and tend to exhibit a double peaked structure. Since \nthp\ is an optically thin midplane ionization tracer, this discrepancy may be indicative of cosmic ray exclusion by magnetic shearing within the disk itself \citep{fujii_2022}. None of the models reproduce the observed emission plateau and edge. We discuss interpretations of the modeling results for the \nthp \ $J = 3 - 2$ emission morphology in depth in Section \ref{sec:disc-morp-nthp}. Overall, SSM-F emerges as the best global fit for both \nthp \ $J = 3 - 2$ ($\chi^{2}_{red}$ = 23.7) and \nthp \ $J = 4 - 3$ ($\chi^{2}_{red}$ = 1.48). 

As shown in the ``Mean" panel of Figure \ref{fig:chisq_all}, SSM-F is the best global fit ($\chi^{2}_{red, mean}$ = 13.8) when we consider all four lines weighted equally. The observed emission profiles for all four lines are best fit by a model with a hard X-ray spectrum and a CR ionization rate of 1.1 $\times$ 10$^{-18}$ s$^{-1}$.

 \begin{figure*}
    \centering
    \includegraphics[scale=0.65]{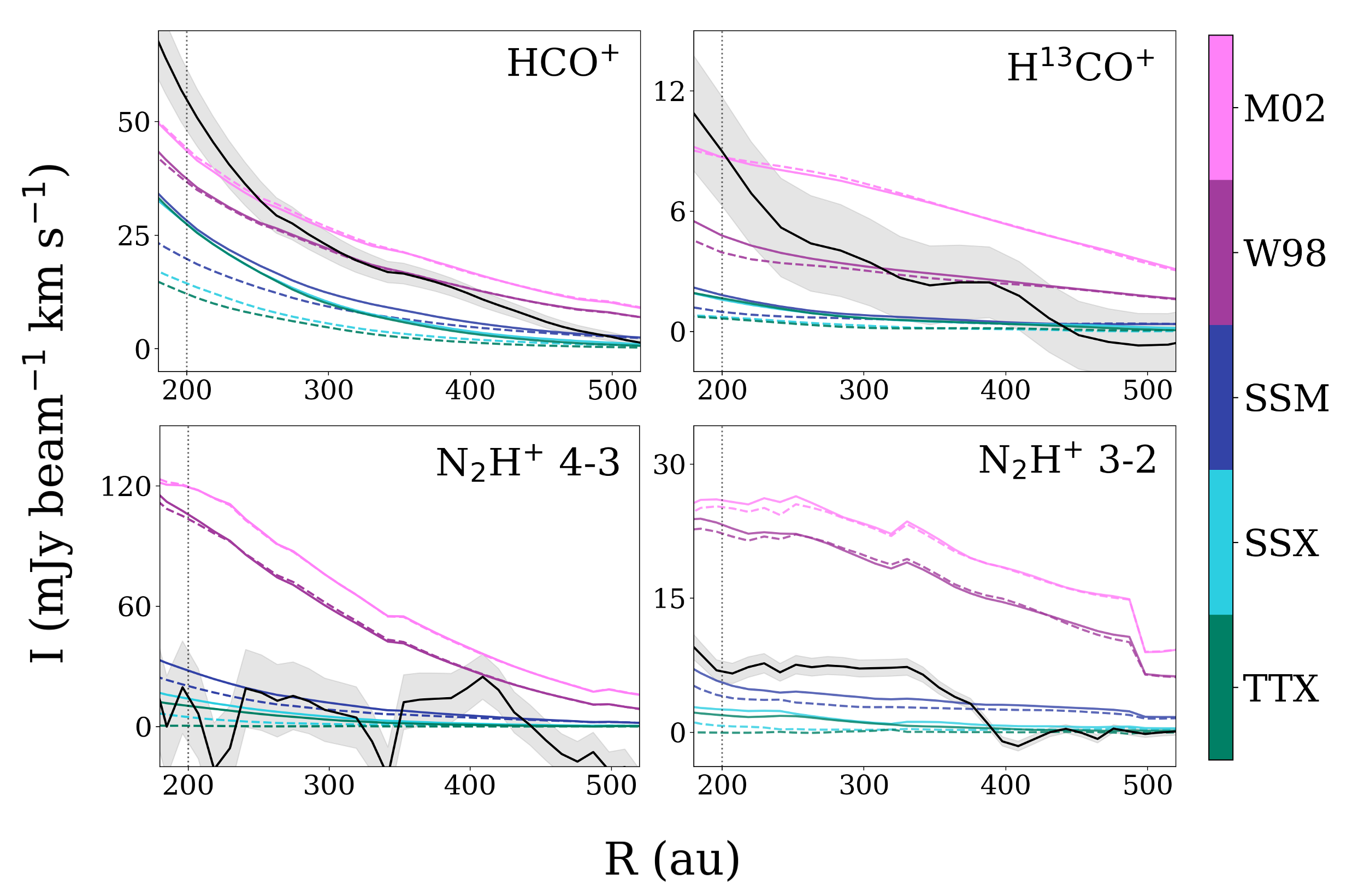}
    \caption{\normalsize{Observed and model radial profiles for the outer disk (R $>$ 200 au). In each panel the solid black line is the observed emission profile and the associated light gray shading represents the uncertainties on the profile from \texttt{gofish} plus the nominal 10\% flux uncertainty from ALMA. The dashed vertical line marks 200 au, which we consider to be the transition radius from inner to outer disk. The five different CR models are color coded according to the colorbar and labels on the right. For the models, dashed lines denote a quiescent XR spectrum while solid lines denote a hardened XR spectrum.}}
  \label{fig:radprofs_out}\vspace{-1mm}
\end{figure*}

\subsubsection{High Turbulence Models}\label{sec:res-glob-hi}

Given the discovery of strong turbulence in this disk by \citet{flaherty}, we examine how including high turbulence into the broadening factor (b = 0.5c$_{s}$) in the radiative transfer modeling impacts our observed line emission. For lines that tend to be more optically thin, we see less of a change, while for optically thick lines more emission becomes ``visible" in the image plane, and thus the velocity-integrated fluxes tend to be higher. 

Increasing the turbulent velocities improves the mean $\chi^{2}_{red}$ scores by 4$\%$ across the full disk and 3$\%$ in the outer disk. The resulting best-fit model for each line remains unchanged aside from \nthp\ $J = 4 - 3$, which changes from SSM-F to SSX-F. The $\chi^{2}_{red}$ score for the best fit model for all four lines is slightly improved from 13.8 to 12.3. Thus we consider the high turbulence SSM-F model to be our best fit, however we do not extend this claim to suggest that this difference is enough to use these data as evidence of high turbulence alone. We discuss the nuances of this result further in Section \ref{sec:disc-ion-env}.

\subsection{Outer Disk Fitting}\label{sec:res-out}

Examination of the radial profile morphology reveals a clear delineation between the inner and outer disk. The inner disk emission tends to be fit best by models producing a high rate of ionization by both CRs and X-rays, while the more diffuse ionization in the outer disk is best reproduced by models with reduced rates of CR ionization. All of the models for \hcop \ and \htcop \ are under-bright in the inner disk, possibly indicating that there are additional physical or chemical processes our model is not encapsulating, especially within the large gap of this disk. Moreover, our models suggest that emission from these molecules is optically thick and is not necessarily tracing the rate of ionization but rather the temperature of those regions. We perform outer disk fitting in order to isolate the regions in which these molecules may offer constraints on the rate of ionization in the outer disk (Figure \ref{fig:radprofs_out}).

We consider the outer disk to be radii $>$ 200 au, as this is the radius at which \hcop \ and \htcop \ are no longer universally under-bright. Additionally, 200 au is the radius at which \nthp \ $J = 3 - 2$ emission ``swaps" from preferring the high CR models to low CR models. In terms of disk structure, the dust disk is smooth beyond the gap at 111 au \citep{hashimoto_2021} so we don't expect any complexities due to substructure to have an effect exterior to 111 au. Thus 200 au is a reasonable choice for representing the shift in chemistry and physics from inner disk to outer disk.

There are some notable changes in our $\chi^{2}_{red}$ fitting results when we consider only the outer disk (Figure \ref{fig:chisq_all}). \hcop\ shifts to favor an even higher CR rate (M02), while \htcop\ sees an increase in agreement for the lower CR models and a relative decrease in agreement for M02. Extremely reduced CR rates (SSX and TTX) are further ruled out for \nthp\ $J = 3 - 2$ while extremely high CR rates are further ruled out for \nthp\ $J = 4 - 3$. The best-fit model for all four lines is still SSM-F, for both the standard ($\chi^{2}_{red}$ = 17.3) and high turbulence ($\chi^{2}_{red}$ = 12.8) models.

\section{Discussion}\label{sec:disc}

\subsection{The Ionization Environment of DM Tau}\label{sec:disc-ion-env}

Our modeling of \hcop, \htcop, and \nthp\ suggests that a high rate of ionization is required to reproduce bright emission of molecular ions in the inner disk, while a lower rate of ionization by CRs is required to fit the outer disk where the observed emission from all four lines is much more diffuse. This result is consistent with a strongly ionized inner disk followed by a steep ionization gradient and a low ionization rate in the outer disk. Previous studies that suggest the presence of CR gradients in TW Hya and IM Lup \citep{cleeves15_twhya, Seifert_2021} found the opposite trend of $\zeta_{CR}$ suppression in the inner disk and, in the case of IM Lup, an increase in $\zeta_{CR}$ towards the outer disk. TW Hya is close to DM Tau in age \citep[$\sim 3-10$ Myr;][]{vacca_2011} and disk mass \citep[0.05 \solarmass;][]{bergin_2013} while IM Lup is both younger \citep[1 Myr;][]{mawet_2012} and more massive \citep[$\sim 0.1$ \solarmass;][]{pinte_2008}. While both TW Hya and IM Lup exhibit substructure in their continuum emission \citep{Huang_2018}, neither of these disks have a large inner cavity like DM Tau's 20 au gap.  

The suppression of CR ionization in TW Hya and IM Lup can be explained by the presence of a disk wind blocking external ionizing radiation from reaching the inner disk. It is less straightforward to explain a CR gradient in the case of DM Tau, as we would not expect there to be a ``barrier" blocking incident external CRs from reaching the outer disk while allowing them to ionize the inner disk. Rather than a reduced rate of ionization in the inner disk like IM Lup or TW Hya, DM Tau appears to have some kind of ionization enhancement localized to the inner $\sim$ 150 au. 

In the outer disk, if one compares the observed \htcop\ and \hcop\ to the models without considering the \nthp, the model with an ISM-like CR rate ($\zeta_{CR}$ $\sim 10^{-17}$~s$^{-1}$) appears to fit reasonably well; however, changes to the CO depletion profile impact the fit to a similar degree (Section \ref{sec:coprof}), and also within the uncertainty of the more optically thin \htcop\ outer disk profile. Given that \htcop\ and \nthp\ trace different vertical regions of the outer disk, if this tension holds with deeper observations, perhaps the data are pointing to a steeper vertical gradient than what is captured in the present models. One possible theory is cosmic ray exclusion within the disk itself by magnetic shearing \citep{fujii_2022}, where cosmic rays are significantly detoured on their way to the disk midplane prior to reaching the \nthp\ layer. The sensitivity of the data is not sufficient to definitively say, but it provides an interesting avenue for future follow up.

Both DM Tau and IM Lup exhibit evidence of turbulence in the upper layers of the outer regions of their disks while TW Hya does not \citep{flaherty,flaherty_24}. While DM Tau and IM Lup each have unique ionization structures, both disks have higher rates of CR ionization ($\gtrsim 10^{-18}$ s$^{-1}$) than TW Hya ($\lesssim 10^{-19}$ s$^{-1}$). Our results alone do not suggest definitively that DM Tau's turbulence is MRI-driven. However, we do find that the inner disk is particularly well-ionized compared to the outer disk. The relatively high degree of ionization in the inner disk may be related to the large 20 au dust gap or another mechanism not captured by our modeling efforts. While we cannot identify the exact mechanism in this work, it is clear that ionizing radiation from the central star plays an important role in the inner disk and could play a role in driving DM Tau's previously discovered turbulence. Robust constraints on disk turbulence and ionization in larger samples will be critical for understanding the role of ionization in facilitating the redistribution of disk material via the MRI.

\begin{figure}
    \centering
    \includegraphics[width=1.0\columnwidth]{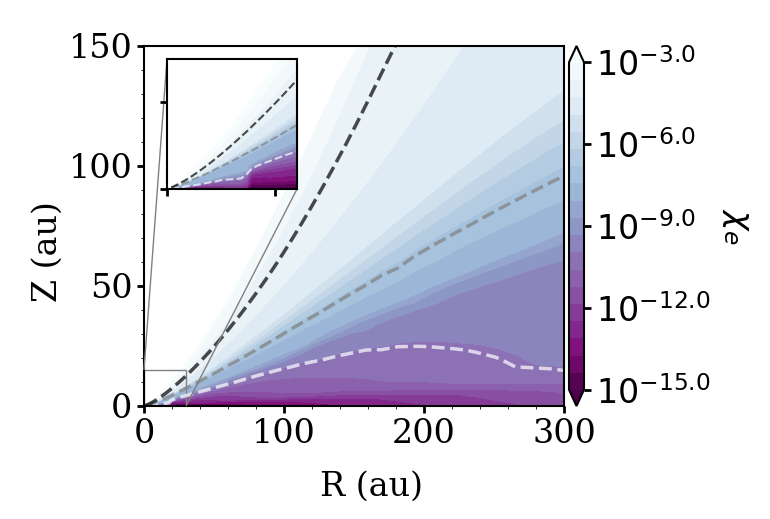}
    \caption{\normalsize{2D ionization fraction (shown as the fraction of electrons, $\chi_{e}$) for our best-fit model, SSM-F. The black, grey, and white contours represent ionization fractions of 1 $\times$ 10$^{-4}$, 1 $\times$ 10$^{-7}$, and 1 $\times$ 10$^{-10}$, respectively. }}
  \label{fig:ion-frac}\vspace{-1mm}
\end{figure}

Our global best fit model, SSM-F, is an intermediate $\zeta_{CR}$ model with a flaring X-ray state. This model comes closest to reproducing the bright emission from molecular ions in the inner disk without overproducing emission from molecular ions in the outer disk. Figure \ref{fig:ion-frac} shows the 2D ionization fraction for the SSM-F model. Our best-fit model suggests that the ionization fraction in DM Tau varies from $\chi_{e}$ $>$ 1 $\times$ 10$^{-4}$ at the surface to $\chi_{e}$ $<$ 1 $\times$ 10$^{-10}$ in the midplane, in agreement with previous estimates of DM Tau's ionization fraction \citep{oberg_2011,Teague_2015}. The inner disk emission from molecular ions in DM Tau is difficult to reproduce using our model under ``normal" assumptions. Even our highest $\zeta_{CR}$ models with hardened X-ray spectra are unable to reproduce the bright inner disk emission from \htcop\ and \nthp- both optically thin tracers. All of this suggests that our models fail to capture some aspects of the chemistry and physics that may be enhancing ionization in DM Tau's inner disk.  

In our models we vary cosmic ray ionization and test the effects of X-ray variability by including both quiescent and hardened X-ray spectra. In reality there may be more complex physics at play. It is possible that the deficit of material within 20 au allows the disk beyond 20 au to be more readily ionized by X-rays, since the attenuation along lines of sight to the star is less than for a full disk. Additionally, ionization by X-rays impacting the inner disk could be non-uniform due shadowing from an asymmetric and/or misaligned inner dust disk. High resolution dust continuum images from \citet{hashimoto_2021} show two asymmetries in the 20 au dust ring, and suggest that the inner ring at 3 au could be misaligned from the outer ring at 20 au, though their visibility analyses find that the rings' PA and inclination values agree within 3$\sigma$.

Another possible cause of an elevated inner disk ionization state could be additional ionization from stellar energetic particles (SPs). Modeling by \citet{rab_2017} suggests that while SPs cannot penetrate the disk midplane, they can dominate ionization in the warm molecular layer for regions close to the star. As a result of additional ionization by SPs, \hcop\ and \nthp\ column densities may be elevated by a factor of $\mathbf{\sim 3-10}$ in the inner disk ($\leq$ 200 au), especially for models with low CR ionization rates. The SP ionization rate for the active T Tauri star model in \citet{rab_2017} is  $\zeta_{SP}$ $\sim 10^{-13}-10^{-12}$~s$^{-1}$ at a vertical column of 10$^{21}$ cm$^{-2}$. This rate is  significantly higher than both the quiescent and ``flaring" XR ionization rates in our models ($\zeta_{XR}$ $\sim 10^{-15}-10^{-14}$~s$^{-1}$) for the same gas column at 10 au. This suggests that SPs could be a competitive source of ionization in DM Tau's inner disk, possibly explaining the elevated emission in \hcop, \htcop, and \nthp\ that we observe. Additionally, \citet{cabedo_2023} find enhanced ionization at small radii that may be explained by cosmic ray acceleration close to the B335 protostar. Both of these results suggest that SPs and CR acceleration close to the central star could play a role in boosting inner disk ionization.

\subsection{\nthp \ Radial Structure }\label{sec:disc-morp-nthp}

Since \nthp \ forms efficiently in the absence of gas-phase CO, observations of this molecule have been used to probe the CO snowline \citep{qi_2013}. The emission peak of \nthp \ $J = 3 - 2$ has previously been interpreted as a direct tracer of CO freeze-out in DM Tau \citep{qi_2019}, however our models struggle to reproduce the location of the peak \nthp \ $J = 3 - 2$ emission at 100 au (Figure \ref{fig:radprofs_glob}, lower right panel). Low CR models (SSM, SSX, TTX) produce peak emission closer to 135 au. High CR models (M02, W98) produce two ``peaks", at $\sim$ 55 au and $\sim$ 170-180 au. Model \nthp \ emission is enhanced by an X-ray flare if present, except for the inner disk emission for M02 models which follows the opposite behavior. No models reproduce the emission plateau and subsequent drop-off between 300 and 400 au. While the SSM-F model emerges as the best global fit for this line ($\chi^{2}$ = 23.7), it does not reproduce the observed emission peak. An examination of the radial emission morphology for this line offers several other possible interpretations. 

If we follow the interpretation of the \nthp \ emission peak as a direct tracer of the CO snowline in DM Tau, then it could be the case that our models have too much gas-phase CO in the inner disk leading to the underproduction of \nthp \ at 100 au. \citet{Zhang_2019} found that the level of CO depletion in DM Tau ranges from a factor of $\sim$ 10 in the inner disk to a factor of $\sim$ 1 in the outer disk. We use an ``average" (factor of 5) in our models, but it's possible that a more detailed model with radial CO depletion is necessary to reproduce the \nthp \ emission morphology, particularly in the inner disk where CO is most depleted. 

Our average-depleted model yields an C$^{18}$O integrated flux of 4.9 Jy km s$^{-1}$, which agrees to within a factor of 2 of the 3.0 Jy km s$^{-1}$ C$^{18}$O flux reported by \citet{Zhang_2019}. The difference in flux is likely due to the fact that our gas model has a smoothly continuous inner disk. Our over prediction of C$^{18}$O in the inner disk is consistent with a more radiation-permeable inner disk, which could also increase the \hcop\ abundance in this inner region. Detailed modeling of the impact of the inner disk gas structure is left to future work as it requires higher resolution molecular ion data of the inner disk. 

Since \nthp\ is also a key constraint on ionization in the outer disk, and thus motivates our findings of a reduced $\zeta_{CR}$, we test the robustness of our \nthp\ $J = 3 - 2$ fits with varying CO depletion. We compared the radial profile results from our ``average" (factor of 5) depletion model to those from models with no CO depletion and a factor of 10 CO depletion. The resulting \nthp\ $J = 3 - 2$ radial profiles are relatively insensitive to the level of CO depletion (varying by a factor of $<$ 2 in flux) and the effect of CRs dominates the effect of CO depletion in the outer disk. Thus, even if we leave CO undepleted, a reduced rate of ionization is required to fit the observations. 

Another possibility discussed in Section \ref{sec:disc-ion-env} is the presence of a radial gradient in CRs. It is clear from the models that the \nthp\ distribution observed in the disk is very sensitive to the assumptions about CR and X-ray fluxes. However, if CRs are responsible, the outer disk provides strong constraints on the absence of CR ionization. Thus some other mechanism causing additional ionization close to the star or attenuation of CRs in the outer disk is required that is not taken into account in our models. 

Finally, another mechanism that could reproduce the \nthp\ peak is a change in the model's midplane temperature. Our temperature model was derived from observations of CO, which do not explicitly probe the midplane. However, using the \texttt{RADEX} radiative transfer code \citep{radex}, the ratio of observed fluxes between the $J = 4 - 3$ and $J = 3 - 2$ transitions at 100 au suggests a temperature of 25 K. This agrees with our model temperatures at 100 au, which range between 20-30 K from midplane to surface. \citet{qi_2019} are able to reproduce the \nthp\ $J = 3 - 2$ radial profile by including a Vertically Isothermal Region above the Midplane (VIRaM layer) in their temperature model. While our CO-derived temperature structure appears to produce reasonable midplane temperatures based on the \nthp\ line ratio, future work with stronger midplane temperature constraints and detailed features such as a VIRaM layer could improve the fit for the \nthp\ emission peak and offer additional insight into midplane ionization.

Ultimately, the emission morphology of \nthp \ is dependent on a combination of effects from ionization, temperature, and gas-phase CO abundance. These parameters/processes are degenerate with one another, and additional modeling efforts are necessary to disentangle their effects. While we do not explore this degeneracy fully in this work, such efforts will help to determine the usefulness of \nthp \ as a direct observational tracer of the CO snowline as well as its usefulness for constraining midplane ionization in a larger sample.

   \begin{figure}
    \centering
    \includegraphics[width=1.0\columnwidth]{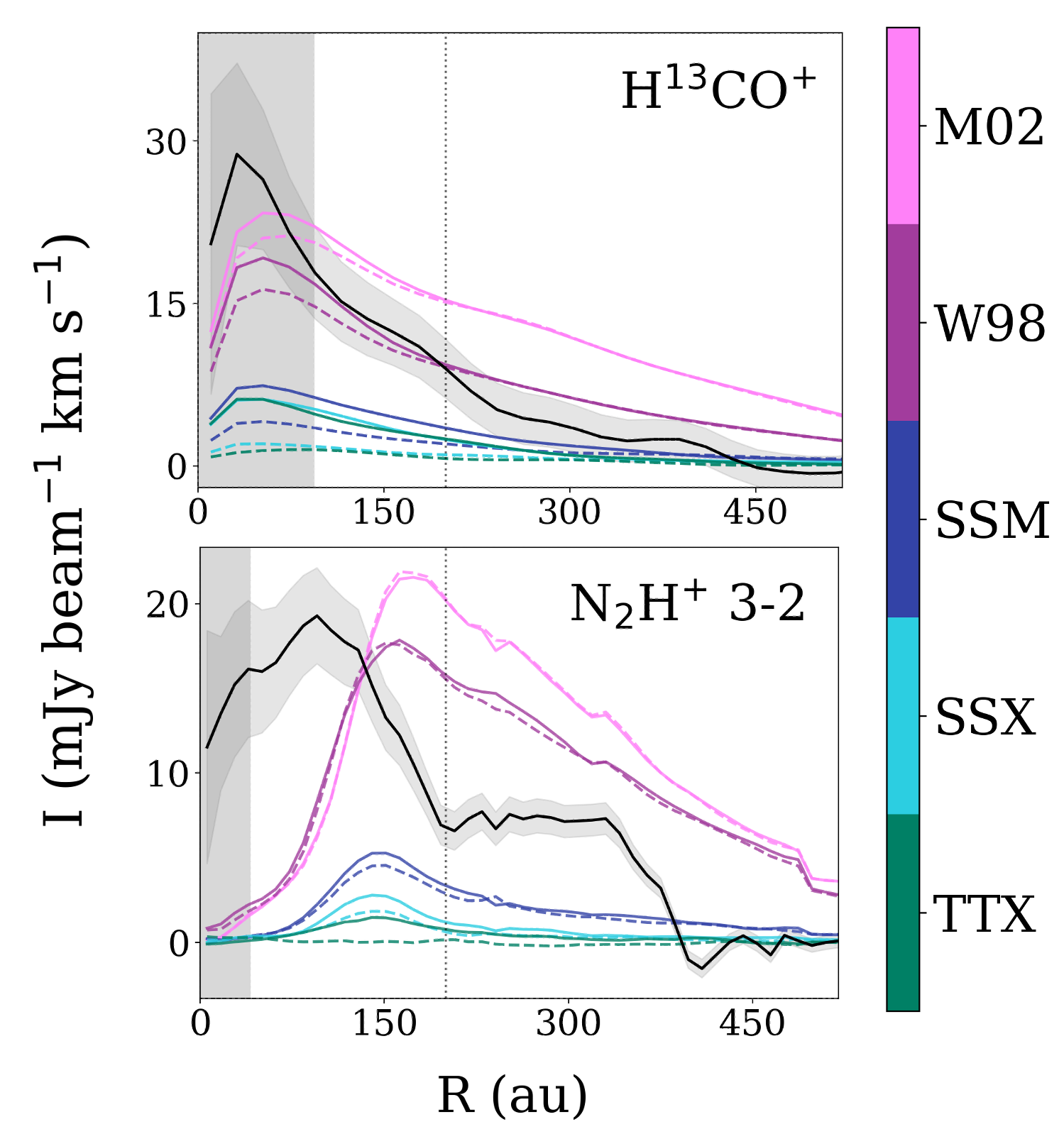}
    \caption{\normalsize{Observed and model radial profiles of \htcop\ and \nthp\ for the models with no CO depletion. The five different CR models are color coded according to the colorbar and labels on the right while the observed profiles are plotted in solid black. Dashed lines denote models with a quiescent XR spectrum while solid lines denote a hardened XR spectrum.}}
  \label{fig:fullco}
\end{figure}

\subsection{Sensitivity to CO Depletion Profile}\label{sec:coprof}

We tested the sensitivity of our models to the CO depletion profile by running an additional model grid with no CO depletion (initial CO abundance = 1.3 $\times$ 10$^{-4}$). As discussed above in Section \ref{sec:disc-morp-nthp}, one possible explanation for the lack of inner disk emission from \nthp\ in our models was an over-abundance of gas-phase CO (a factor of 5 depletion compared to the factor of 10 depletion inferred by \citet{Zhang_2019}). However, decreasing the amount of CO in the inner disk also leads to a decrease in \hcop\ and \htcop\ which contradicts what we observe. As shown in Figure \ref{fig:fullco}, increasing the initial CO abundance for the model grid results in brighter \htcop\ and dimmer \nthp\ in the inner disk. This aligns with our expectations, as \htcop\ forms readily in the presence of CO while \nthp\ is destroyed. Both species are sensitive to the degree of CO depletion, but in opposite ways.

Since brighter inner disk fluxes are required to fit the observed emission for both species, the CO depletion profile itself cannot account for the ``boost" in emission in the inner disk. In the outer disk, where we expect there to be little to no CO depletion \citep{Zhang_2018}, high CR models can be ruled out for both \htcop\ and \nthp, further motivating the presence of CR shielding across the full disk extent. In reality DM Tau likely exhibits a complex CO depletion profile (higher depletion possibly increasing \nthp\ production in the inner disk) as well as additional physical processes leading to an increase in ionizing radiation, which would account for the bright inner disk emission from both species.

\subsection{Uses of Different Ionization Tracers}\label{sec:disc-use}

By forward-modeling the DM Tau disk in detail we are not only able to gain valuable insights to the ionization environment of this system, but also examine the usefulness of each of the ionization tracers included in this work. \hcop \ $J = 4 - 3$ is considered to be an optically thick tracer while \htcop \ $J = 3 - 2$ is expected to be optically thin. In DM Tau we see bright \hcop \ emission coming from the high-velocity wings, suggesting that there is a no significant depletion of gas in the inner disk despite the large 20 au dust cavity. In more compact sources it is possible that \hcop\ could be optically thin in the inner disk. Thus, this tracer could be very useful for constraining inner disk conditions for compact sources. In bright sources, it is clear that \htcop\ is a necessary complementary line to observe in addition to optically thick \hcop.

Models and observations of \hcop \ suggest that its abundance in the disk surface can be altered by rapid X-ray changes from the central star, such as X-ray flares \citep{waggoner_2022, waggoner_2023}. All of our DM Tau models with a hardened ``flaring" XR spectrum have brighter peaks than their quiescent counterparts, though the disparity is wider for low CR models. Our models suggest that in the presence of a high $\zeta_{CR}$ (M02, W98), the effect of an X-ray flare on the emission of optically thick \hcop \ is subtle, only changing the peak profile intensity ($\Delta$I$_{peak}$) by an average of 5 mJy beam$^{-1}$ km s$^{-1}$. On the other hand, an X-ray flare under low $\zeta_{CR}$ conditions (SSM, SSX, TTX) will have a more noticeable effect on the observed \hcop \ emission, with $\Delta$I$_{peak}$ = 50--80 mJy beam$^{-1}$ km s$^{-1}$. Thus if one wants to observe the effects of X-ray flares using \hcop, sources with low $\zeta_{CR}$ may be favorable targets. 

As discussed in Section \ref{sec:disc-morp-nthp}, there are several interpretations of our modeling results for \nthp\ and it is not fully clear whether or not \nthp\ is a direct tracer of the CO snow line in DM Tau. Larger samples with sensitive and resolved observations of additional \nthp\ transitions and/or other optically thin ionization tracers which form in the presence of gas-phase CO (e.g., \htcop, \dcop) will help to guide modeling efforts to clarify the role of \nthp\ as a direct tracer of CO freeze out. Overall, \nthp\ is an effective cold ionization tracer that is sensitive to both CRs and X-rays. \nthp\ may be particularly enhanced in disks with substructure that allows for the penetration of X-rays deeper into the disk midplane. 

It is important to use multiple tracers that probe different disk emitting heights to interpret the ionization environments of protoplanetary disks. DM Tau exhibits evidence of both radial and vertical ionization gradients. In our fitting of the radial profiles, a boost in inner disk ionization is required for all three molecular ions even for some of the highest CR models. When we consider the different emitting heights of our optically thin tracers \htcop\ and \nthp, their emission profiles suggest a vertical gradient in CR ionization that may not be explained by attenuation alone. Overall, the small sample of disks with well-characterized ionization environments-- DM Tau, TW Hya, and IM Lup-- exhibit a variety of ionization structures which appear to be impacted by different shielding mechanisms at different disk locations. An exploration of both radial and vertical ionization gradients is required in order to differentiate between the many factors governing disks' complex ionization environments. Thus, studies of disk ionization should include resolved observations of at least two optically thin tracers tracing distinct disk emitting heights.

\section{Conclusions}\label{sec:conc}

We present new observations of \hcop\ $J = 4 - 3$, \htcop\ $J = 3 - 2$, and \nthp\ $J = 4 - 3$ in the DM Tau protoplanetary disk. Using these observations along with archival observations of \nthp\ $J = 3 - 2$, we aim to forward model DM Tau's ionization environment. Our main findings are as follows:

\begin{itemize}
\itemsep0em
    \item When considering all three molecular ions, DM Tau's ionization state is best fit by a model with a CR ionization rate an order of magnitude lower than that of the ISM ($\zeta_{CR}$ = 1.1 $\times$ 10$^{-18}$ s$^{-1}$) and a hardened X-ray spectrum, post-processed with high turbulent velocities (b = 0.5c$_{s}$). The improvement with high turbulent velocity is not significant, but leads to slightly better fits. The ionization fraction in our best-fit model varies from $\chi_{e}$ $\gtrsim$ 1 $\times$ 10$^{-4}$ at the surface to $\chi_{e}$ $\lesssim$ 1 $\times$ 10$^{-10}$ in the midplane.
    \item DM Tau exhibits a steep drop in ionization from the inner disk to the outer disk that our models struggle to reproduce with one ionization rate $\zeta_{CR}$. We speculate that disk has some shielding mechanism reducing the incident CR rate across the full 500 au extent, but that there are either (1) additional physical effects not modelled, like stellar energetic particles, or (2) the presence of substructure making the inner disk more permeable to CRs. Both of these scenarios deserve dedicated modeling efforts. 
    \item Our observations appear to trace a flaring state of DM Tau (HR = 0.3). In our fitting of individual lines, all four are best fit by models with the hardened X-ray spectrum. Additional observations of \hcop, \htcop, and \nthp\ in DM Tau at different epochs could reveal evidence of X-ray variability.

\end{itemize}

This study provides new constraints on DM Tau's ionization environment through detailed 2D forward modeling. Our methods allow us to examine the roles of different sources of ionization, leading to a better understanding of the physical and chemical conditions governing disk evolution and planet formation. This work highlights the need for multi-line studies which allow us to probe different vertical layers of the disk. By combining molecules that trace different layers and by examining the spatially resolved radial profiles, we are able to rule out a high CR ionization rate across the full radial and vertical disk extent and identify a possible enhancement occurring within the inner disk of DM Tau. In general, constraining ionization in protoplanetary disks remains challenging because 1) ionization is provided by a diverse set of sources, 2) the efficacy of those sources varies throughout the disk, and 3) the sources vary from disk to disk. High resolution, multi-line studies of ionization tracers in large samples will be critical for disentangling the complexity of disk ionization and its sources.


\section*{Acknowledgments}

We thank the anonymous referee for their insightful comments that improved the quality of this manuscript. DEL thanks Z. Y. Li and R. Loomis for helpful discussions about the results presented in this paper. This paper makes use of the following ALMA data: ADS/JAO.ALMA$\#$2019.1.00379.S
and ADS/JAO.ALMA$\#$2015.1.00678.S. ALMA is a
partnership of ESO (representing its member states),
NSF (USA) and NINS (Japan), together with NRC
(Canada), MOST and ASIAA (Taiwan), and KASI (Republic of Korea), in cooperation with the Republic of
Chile. The Joint ALMA Observatory is operated by
ESO, AUI/NRAO and NAOJ. The National Radio Astronomy Observatory is a facility of the National Science Foundation operated under cooperative agreement by Associated Universities, Inc. We are grateful to the Atacameño (Likan Antai) Elders-- the first astronomers of Atacama-- whose wisdom and knowledge we continue to learn and benefit from as ALMA users. DEL and LIC acknowledge support from NASA ATP 80NSSC20K0529. DEL also acknowledges support from the Virginia Space Grant Consortium. LIC also acknowledges support from NSF grant no. AST-2205698, the David and Lucille Packard Foundation, and the Research Corporation for Scientific Advancement Cottrell Scholar Award. V.V.G. gratefully acknowledges support from FONDECYT Regular 1221352, and ANID CATA-BASAL project FB210003. The modeling conducted in this paper was carried out on the University of Virginia’s Rivanna High Performance Computing Cluster.

%

\vspace{5mm}
\facilities{ALMA}
\software{CASA \citep{casa_mcmullin_2007}, LIME \citep{brinch_2010}, vis\textunderscore sample \citep{Loomis_2018}, bettermoments \citep{teague_2018_bm}, GoFish \citep{GoFish}, keplerian\textunderscore mask \citep{teague_2020_kep}.}





\appendix
\section{Azimuthally averaged optical depth profiles}

\begin{figure*}[h]
    \centering
    \includegraphics[scale=0.44]{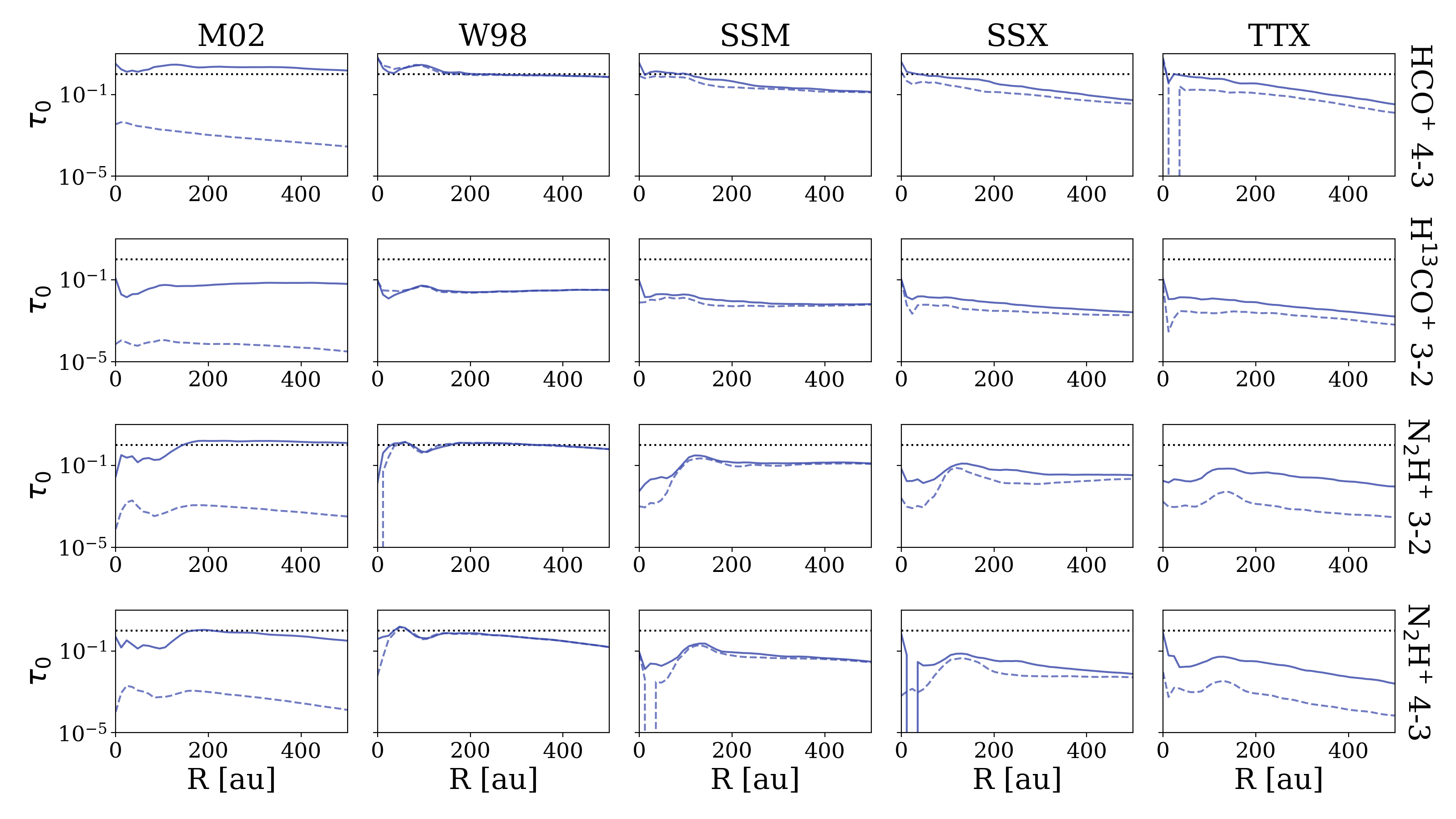}
        \caption{\normalsize{Model optical depth profiles produced with LIME. The dotted black line shows the $\tau$ = 1 surface. The column labels (top) denote the five different CR models and the row labels (right) denote the four different molecular lines. Dashed lines represent models with a quiescent XR spectrum and solid lines represent models with a hardened XR spectrum.}}
  \label{fig:tau_profs}\vspace{-1mm}
\end{figure*}

Figure \ref{fig:tau_profs} shows the azimuthally averaged optical depth profiles of \hcop\ $J = 4 - 3$, \htcop \ $J = 3 - 2$, \nthp\ $J = 4 - 3$, and \nthp\ $J = 3 - 2$ for the grid of models described in Section \ref{sec:mod}. 

\end{document}